\documentstyle[12pt]{article}
\newcommand {\gl}{\gamma_L}
\newcommand {\gs}{\gamma_S}

\newcommand {\dm}{\Delta m}
\newcommand {\dl}{\Delta \lambda}

\newcommand {\el}{\epsilon_L}
\newcommand {\es}{\epsilon_S}

\newcommand {\Rel}{{\rm Re}\epsilon_L}

\newcommand {\Res}{{\rm Re}\epsilon_S}

\newcommand {\Rep}{{\rm Re}\epsilon}

\newcommand {\rss}{\rho_{SS}}
\newcommand {\rll}{\rho_{LL}}
\newcommand {\rsl}{\rho_{SL}}

\newcommand {\xtt}{x_{33}}
\newcommand {\xdt}{x_{23}}

\newcommand {\El}{e^{-\bar{\gamma}_L t}}
\newcommand {\Es}{e^{-\bar{\gamma}_S t}}
\newcommand {\E}{e^{-\frac{1}{2} \gamma t}}

\begin{document}
\newcommand{\ECM}{\em Departament d'Estructura i Constituents de la
Mat\`eria
                  \\ Facultat de F\'\i sica, Universitat de Barcelona \\
                     Diagonal 647, E-08028 Barcelona, Spain \\
                                      and           \\
                                    I. F. A. E.}

\def\thefootnote{\fnsymbol{footnote}}
\pagestyle{empty}
{\hfill \parbox{6cm}{\begin{center}
                                    September 2000
                     \end{center}}}
\vspace{1.5cm}

\begin{center}
\large{Can CP entanglement with the environment mask CP violation?}

\vskip .6truein
\centerline {J. Taron\footnote{e-mail: taron@ecm.ub.es}}
\end{center}
\vspace{.3cm}
\begin{center}
\ECM
\end{center}
\vspace{1.5cm}

\centerline{\bf Abstract}
\medskip
We consider departures from hamiltonian dynamics in the evolution of neutral
kaons due to their interactions with the environment that generate 
entanglement among them.
\vspace{2cm}

\begin{center}
{\it Talk presented at the Conference QUARKS2000, Pushkin, Saint Petersburg.}
\end{center}

\newpage
\pagestyle{plain}

\section{Introduction}
CP violation was first measured in \cite{CCFT} long ago, 
in the mixing of the neutral 
kaons, the so called {\it indirect} CP violation, 
whereas it was not until recently that {\it direct}
violation, in the decay amplitudes to two pions, was
experimentally established (see \cite{BS} and references therein).
There are at present a few dedicated experiments
in project to make precision measurements of the CP breaking parameters.
The analysis of their data may require to take properly into account the
effects of {\it decoherence} due to entanglement of the neutral kaon
system with the environment in which it evolves, which entails a pure
kaon state to evolve into a mixed one. 
We consider the evolution
of neutral kaons in the presence of matter, in an environment that is not the
perfect vacuum, which generates departures from hamiltonian dynamics
in the kaon sub-system. These are effects that exist in addition to the
weak interactions and are dominated by the strong interactions of the kaons
and the environment: they lead to and effective breaking of charge
conjugation, because the environment is made of matter, not of anti-matter;
and to an effective breaking of time-reversal invariance associated to the
large number of degrees of freedom in the environment that leads to 
irreversibility.

Similar analysis can be found in the literature 
\cite{BF}, \cite{PH}, \cite{EMN}. There, the 
motivation is to study decoherence effects that come from quantum gravity.
Ours is the interaction with the environment that is not the vacuum.

The neutral kaon evolution in the vacuum is analyzed with the Schr\"odinger
equation
\begin{equation}
i \dot{\rho}(t)= H \rho(t) - \rho(t) H^\dagger,
\label{schrodinger}
\end{equation}
where $H$ is 
the LOY \cite{loy} 
effective hamiltonian $H=\sum_{I=S,L} \lambda_I 
|K\rangle \langle \tilde{K}_I|$, which is not hermitean and its eigenstates
$\lambda_I= m_I - \frac{i}{2} \gamma_I$ possess a non-vanishing imaginary
part of their decay rates. $S,L$ stand for the {\it short} and {\it long}
components, and $|\tilde{K}_I \rangle$ are such that 
$\langle \tilde{K}_I | K_J \rangle = \delta_{IJ}$. 
Recall that $m_S \simeq m_L \simeq 500 \; MeV$, 
$\Delta m = m_L-m_S =3.5 \times 10^{-12} \; MeV$, 
$\gamma_S=7.3\times 10^{-12} \; MeV$ and 
$\gamma_S=1.3 \times 10^{-14} \; MeV$. 

The $|K_I\rangle$ have
diagonal evolution $|K_I(t)\rangle= e^{-i \lambda_I t}|K_I(0)\rangle $. In
terms of the CP eigenstates
\begin{equation}
|K_S\rangle=\frac{1}{\sqrt{1+|\epsilon_S|^2}}\left( |K_1\rangle +
\epsilon_S |K_2 \rangle \right), \;\;\;\;\;
|K_L\rangle=\frac{1}{\sqrt{1+|\epsilon_L|^2}}\left( |K_2\rangle +
\epsilon_L |K_1 \rangle \right);
\label{def}
\end{equation}
they reduce to $|K_1 \rangle$, $|K_2 \rangle$, respectively, i.e.,
to the CP eigenstates $\frac{1}{2} \left( |K^0\rangle \pm |\bar{K}^0 \rangle
\right)$ if CP is conserved.

At this point we recall that CPT conservation requires 
$\epsilon_S=\epsilon_L$; T invariance, $\epsilon_S+\epsilon_L=0$;
and CP, $\epsilon_S=\epsilon_L=0$.

Indirect CP violation has been measured 
\begin{equation}
\epsilon \simeq 2.3 \times 10^{-3} e^{i \pi/4},
\label{epsilon}
\end{equation}
(CPT conservation is assumed, $\epsilon_S=\epsilon_L\equiv \epsilon$.)
Direct CP violation has also been measured
\begin{equation}
\epsilon' /\epsilon \sim 2 \times 10^{-3},
\label{epsilonprima}
\end{equation}
where
$$\frac{A(K_L \to \pi^+ \pi^-)}{A(K_S \to \pi^+ \pi^-)} \equiv
\epsilon + \epsilon', \;\;\;\;
\frac{A(K_L \to \pi^0 \pi^0)}{A(K_S \to \pi^0 \pi^0)} \equiv
\epsilon - 2 \epsilon'.$$

\section{Interaction with the environment}
The evolution of the {\it total} system [Kaon + (Large) Environment]
is unitary, with the time evolution operator 
given by $U(t)=\exp(-i H_{Total}t)$;
whereas the dynamics of the kaon sub-system alone is obtained by
tracing out the environment degrees of freedom:
\begin{equation}
\rho_K(t)= Tr_{ENV} \left( U(t) \rho_{K+ENV}(0) U^\dagger (t) \right).
\label{evolution}
\end{equation}
This gives a complicated evolution for $\rho_K$, even if the initial conditions
are a direct product $\rho_{K+ENV}(0)=\rho_K (0) \otimes \rho_{ENV}(0)$.
However, when the interaction is tiny (e.g., as in the case of a 
diluted environment) the dynamics of $\rho_K$ becomes approximately free from
memory effects (Markovian), and it is dictated by a Lindblad master
equation of the form
\begin{equation}
\dot{\rho}=-i \left( H_{eff} \rho(t) - \rho(t) H_{eff}^\dagger \right)
+ \sum_n \left( A_n \rho A_n^\dagger - \frac{1}{2} \rho A_n^\dagger A_n 
- \frac{1}{2} A_n^\dagger A_n \rho \right),
\label{lindblad}
\end{equation}
(henceforth $\rho$ stands for $\rho_K$). Memory effects are being neglected.
Notice that the new piece is linear in $\rho$ but 
{\it quadratic} in the unspecified operators $A_n$: the Lindblad equation
encodes a new dynamics that is not purely hamiltonian. The new term
entails decoherence, it makes pure states evolve into mixed states. Since
the opposite process does not occur, i.e, no mixed state
evolves to a pure one, we demand that the von Neumann entropy
$-\rho \log \rho$ should not increase, under evolution with the new piece
only.

The operators $A_n$ are governed by the strong interactions and the properties
of the medium. Since strong interactions conserve strangeness, which
in the absence of the weak interactions would become a superselection
quantum number, we further demand that {\it any} density matrix, 
at large times, should end up as a mixture of $|K^0 \rangle$
and $|\bar{K}^0 \rangle$. 

The most general parametrization that satisfies the above considerations
leads, for the two state neutral kaon system, to a non-hamiltonian 
contribution in (\ref{lindblad})
\begin{equation}
                 \left(  \begin{array}{cccc}  
                        0 & 0 & 0 & 0  \\
                        0 & s_1 & 0 & 0 \\
                        0 & 0 & s_{22} & s_{23}  \\
                        0 & 0 & s_{23} & s_{33} 
                  \end{array} \right)
                  \left( \begin{array}{c}
                        \rho^0\\ \rho^1 \\ \rho^2 \\ \rho^3
                         \end{array} \right).
\label{coefficients}
\end{equation}
where in the CP eigenbasis $\rho\equiv \frac{1}{2} ( \rho^0 I + 
\vec{\rho} \cdot \vec{\sigma} )$.

The matrix $s_{ij}$ is {\it symmetric}, $s_1, s_{22}, s_{33} >0$, it
has positive eigenvalues the smallest of which is $s_1$. They also
verify the relations \cite{BF}
$$s_1 \leq s_{22}+ s_{33}, \;\;\;\;\; s_{33}^2 \geq (s_1-s_{22})^2,$$
$$s_{22} \leq s_1+ s_{33}, \;\;\;\;\; s_{22}^2 \geq (s_1-s_{33})^2,$$
$$s_{33} \leq s_1+ s_{22}, \;\;\;\;\; s_{1}^2 \geq (s_{22}-s_{33})^2.$$
The hamiltonian part gets a contribution $H_l$ that is of the order 
$o(s_{ij})$.

For example, in an environment of infinitely heavy particles which act
as scattering centers, since no entaglement is generated $s_{ij}=0$
and $H_l$ adopts the Kabir-Good form
\begin{equation}
(H_l)_{Kabir-Good}=\frac{2 \pi n}{m_K} 
\left( f_{K^0}(0) |K^0\rangle \langle K^0| +
       f_{\bar{K}^0}(0) |\bar{K}^0\rangle \langle \bar{K}^0| \right);
\label{kabirgood}
\end{equation}
$f(0)$ is the forward scattering amplitude of the kaon with one center
and $n$ stands for the density of scatterers.

\section{Estimates}
Can these effects be detected? How large do we expect them be?
In order to work out the time evolution of the density matrix
from equation (\ref{lindblad}) we assume $\frac{s_{ij}}{\Delta m}$ of the order
of the smallest observed effect $\epsilon' \sim 10^{-6}$. With the
hierarchy
$$\gamma_L/\gamma_S, \epsilon \sim 10^{-3} >> \delta_K,
\frac{s_{ij}}{\Delta m},
\epsilon', \epsilon^2 \sim 10^{-6},$$
we calculate the evolution up to $o(\frac{s_{ij}}{\Delta m})$. Here, $\delta_K=
\frac{\epsilon_L-\epsilon_S}{2}|_{eff}$ ammounts to effective CPT breaking,
coming from the hamiltonian part of the new piece in (ref{lindblad}).  
On dimensional grounds we take $s \sim \sigma n$, and $\frac{s}{\Delta m}
\sim \frac{\sigma n}{\Delta m}\sim o(\epsilon') \sim 10^{-6}$ can be
achieved with $n \sim 10^{20} /cm^3 \sim [N_{Avogadro}/cm^3]/1000$,
where $\sigma \sim 1 \;fm^2$.

In the basis $|K_S\rangle, |K_L\rangle$ the hamiltonian part of the
evolution is straightforwardly obtained, to all orders in $\epsilon_S,
\epsilon_L$, if needed. With the notation 

$$\bar{\gamma}_L=\gamma_L+\frac{s_{33}}{2}, \;\;
\bar{\gamma}_S=\gamma_S+\frac{s_{33}}{2}, \;\;
\gamma=\gs+\gl+s_1+s_{22};$$ 

$$\xtt=\frac{s_{33}}{2 \gs},\;\;
\xdt=\frac{s_{23}}{2 |\Delta \lambda|}, \;\;
x=\frac{s_{22}-s_1}{2 \Delta m}, \;\;
\phi={\rm Arg} \Delta \lambda.$$. 

$$\dl \equiv \lambda_L - \lambda_S =( m_L - m_S ) + 
\frac{i}{2} (\gamma_S - \gamma_L),$$
the time evolution of an initial $|K^0\rangle$ reads, tto first order in
perturbation theory,
$$\rss(t)=\frac{\xtt}{2} \El- \xdt \cos(\dm t +\phi) \E
+\left( \frac{1}{2} - \Rel +2 (\Rep)^ 2 -\frac{\xtt}{2}+\xdt \cos\phi
\right) \Es$$
$$\rll(t)=\left( \frac{1}{2} - \Res +2 (\Rep)^ 2 +\frac{\xtt}{2}-\xdt \cos\phi
\right) \El + \xdt \cos(\dm t - \phi) \E - \frac{\xtt}{2} \Es $$

\begin{eqnarray}
\rsl(t)&=&\frac{\xdt}{2} e^{-i \phi} \El +
\left[ \left(\frac{1-(\es^{*}+\el)+ 4 (\Rep)^2}{2}+ i\xdt \sin \phi \right)
e^{i \dm t} + \frac{x}{2} \sin(\dm t) \right] \E
\nonumber \\
&-& \frac{\xdt}{2} e^{i \phi} \Es
\label{k0} 
\end{eqnarray}
and similarly for the $|\bar{K}^0\rangle$.

\section{How can the new effects be uncovered?}
The new effects cannot be completely absorbed by redefinitions of the 
parameters in the effective hamiltonian. Except for $\epsilon'/\epsilon$,
in all the other CP violating asymmetries that have been measured
so far associated with the decays into $2 \pi$ and semileptonic into
$\pi^{\pm} l \nu$, they give subleading contributions and are thus expected
to bear little influence in the analysis.

Furthermore,
in the experiments that have been performed, all the measured observables
correspond either to detection of decay rates
at large times, when only the $K_L$
component survives, or to integrated rates. The exception is CPLEAR where
the time evolution can be traced in the interval $\tau_S < t < 20 \tau_S$.
We find that a better determined time evolution could help in
measuring the new parameters of the experiment. It is also found
that it is rather the symmetric combinations of rates that single out the
new coefficients. For example, at large times
$$P(K^0 \to \pi \pi;t) \stackrel{\gamma_S t \gg 1}{\simeq}
\frac{1}{2} (x_{33}+ |\epsilon|^2) e^{-\bar{\gamma_L}t},$$
with an independent measure of $\epsilon$ the parameter 
$x_{33}=\frac{s_{33}}{2 \gamma_S}$ could be determined
-the normalization can be obtained from small t.
It also appears as subleading in
$$P(K^0 \to \bar{K}^0)+P(\bar{K}^0 \to K^0)
\stackrel{\gamma_S t \gg 1}{\simeq} \left(
\frac{1}{2}+ 4 (Re \epsilon)^2 + x_{33} \right) e^{-\bar{\gamma_L}t},$$
from which only an extremely precise experiment could determine it.

Furthermore, mathematically, at short times
$$P(K^0 \to \bar{K}^0) \sim P(\bar{K}^0 \to K^0) 
\stackrel{\gamma_S t \ll 1}{\sim} \frac{1}{2} s_1 t. $$

\section{$\epsilon'/\epsilon$ in two-pion decays}
Recently the double ratio
$$\frac{\Gamma(K_L \to \pi^+ \pi^-)/\Gamma(K_S \to \pi^+ \pi^-)}{\Gamma(K_L \to \pi^0 \pi^0)/\Gamma(K_S \to \pi^0 \pi^0)}$$
has been measured. In a perfect vacuum this is 
$$\frac{|\epsilon + \epsilon|^2}{|\epsilon -2 \epsilon|^2}\sim 1+ 6 Re 
\left( \frac{\epsilon'}{\epsilon} \right).$$

However, in matter the result is 
$$1+6 Re\left( \frac{\epsilon'}{\epsilon_L} \right) \frac{|\epsilon_L|^2}
{|\epsilon_L|^2 + x_{33}}.$$
Since $x_{33}$ is positive, the effect of entanglement with the environment
would decrease the signal, i.e., it would be bigger than measured.
Here $\epsilon_L$ comes from the hamiltonian part of the evolution.
\vspace{1cm}

We believe that a measurement of these parameters in each experiment of
precision would be interesting on its own, in addition to being necessary
for properly subtract them. In this way one would uncover the effects of
entanglement with the environment of the neutral kaons, that certaintly,
unavoidably exist.

\section{Conclusions}
We draw the attention to effects of interaction with the environment in
neutral kaon envolution, that maybe of relevance in the analysis of
the CP breaking parameters, made from more complete and precise data
that will be available soon, in the future (e.g., KLOE in Eurodafne).
The Lindblad master equation provides a {\it universal} form that
incorporates the tiny corrections of non-hamiltonian evolution.
The effects are expected to be more important in B physics for although
the B's decay faster than the kaons, their entanglement with the
environment will be more efficient, given the B meson mass of
about 5 $GeV$, a factor of ten larger than that of kaons.

\vspace*{1cm}
\section{Acknowledgments}
This work was done in a most enjoyable and fruitful collaboration
with Alexander Andrianov and Rolf Tarrach.
The author thanks the organizers of Quarks2000 for such a profitable
conference.
Financial support from CICYT, contract AEN95-0590,
and from CIRIT, contract GRQ93-1047 are acknowledged


\begin{thebibliography}{99}
\bibitem{CCFT} J.H. Christenson, J.W. Cronin, V.L. Fitch and R.Turlay,
Phys. Rev. Lett. 13 (1964) 138.

\bibitem{BS} I.I. Bigi, A.I. Sanda, CP Violation, Cambridge University Press
2000. 

\bibitem{BF} F. Benatti, R. Floreanini, Nucl. Phys. B488 (1997) 335.

\bibitem{PH} P. Huet, M. Peskin, Nucl. Phys. B434 (1995) 3.

\bibitem{EMN} J.Ellis, N.E. Mavromatos, D.V. Nanopoulos, Phys. Lett. B293
(1992) 142.

\bibitem{loy} T.D. Lee, R. Oehme, C.N. Yang, Phys. Rev. 106 (1957) 1340.


\end{thebibliography}
\end{document}